\documentclass[preprint2]{emulateapj}

\newcommand{\degree}{$^{\circ}$}
\newcommand{\um}{$\mu$m}

\shorttitle{25\,$\mu$m imaging of HAeBe}
\shortauthors{Honda et al.}

\begin{document}
\title{High-resolution 25\,$\mu$m imaging of the disks around Herbig Ae/Be stars \altaffilmark{1}}


\author{M. Honda\altaffilmark{2}, K. Maaskant\altaffilmark{3,4},
Y. K. Okamoto\altaffilmark{5}, H. Kataza\altaffilmark{6},
T. Yamashita\altaffilmark{7}, T. Miyata\altaffilmark{8},
S. Sako\altaffilmark{8}, T. Fujiyoshi\altaffilmark{9}, I. Sakon\altaffilmark{10},
H. Fujiwara\altaffilmark{9}, T. Kamizuka\altaffilmark{8}, 
G. D. Mulders\altaffilmark{11},
E. Lopez-Rodriguez\altaffilmark{12}, C. Packham\altaffilmark{12} 
and T. Onaka\altaffilmark{10} 
}


\altaffiltext{1}{Based on data collected at Subaru Telescope, via the time exchange program 
between Subaru and the Gemini Observatory. The Subaru Telescope is operated by the 
National Astronomical Observatory of Japan.}
\altaffiltext{2}{Department of Mathematics and Physics, Kanagawa University,
2946 Tsuchiya, Hiratsuka, Kanagawa 259-1293, Japan}
\altaffiltext{3}{Leiden Observatory, Leiden University, PO Box 9513, 2300 RA Leiden, The Netherlands}
\altaffiltext{4}{Astronomical Institute Anton Pannekoek, University of Amsterdam, P.O. Box 94249, 1090 GE Amsterdam, The Netherlands}
\altaffiltext{5}{Institute of Astrophysics and Planetary Sciences,
Faculty of Science, Ibaraki University, 2-1-1 Bunkyo, Mito, Ibaraki 310-8512, Japan}
\altaffiltext{6}{Department of Infrared Astrophysics, Institute of Space
and Astronautical Science, Japan Aerospace Exploration Agency,
3-1-1 Yoshinodai, Sagamihara, Kanagawa 229-8510, Japan}
\altaffiltext{7}{National Astronomical Observatory of Japan, 2-21-1
Osawa, Mitaka, Tokyo 181-8588, Japan}
\altaffiltext{8}{Institute of Astronomy, School of
Science, University of Tokyo, 2-21-1 Osawa, Mitaka, Tokyo 181-0015, Japan}
\altaffiltext{9}{Subaru Telescope, National Astronomical Observatory of
Japan, 650 North A'ohoku Place, Hilo, Hawaii 96720, U.S.A.}
\altaffiltext{10}{Department of Astronomy, School of Science, University
of Tokyo, Bunkyo-ku, Tokyo 113-0033, Japan}
\altaffiltext{11}{Lunar and Planetary Laboratory, The University of Arizona, Tucson, AZ 85721, USA}
\altaffiltext{12}{Department of Physics \& Astronomy, 
University of Texas at San Antonio, One UTSA Circle, San Antonio, TX 78249, USA}


\begin{abstract}
We imaged circumstellar disks around 22 Herbig Ae/Be stars at 25\,$\mu$m using Subaru/COMICS and Gemini/T-ReCS.
Our sample consists of equal numbers of objects belonging
to the two categories defined by \cite{Meeus2001}; 11
group I (flaring disk) and II (flat disk) sources.
We find that group I sources tend to show more extended emission than group II sources.
Previous studies have shown that the continuous disk is hard to be resolved with 8 meter class telescopes in Q-band due to the strong emission from the unresolved innermost region of the disk. It indicates that the resolved Q-band sources require a hole or gap in the disk material distribution to suppress the contribution from the innermost region of the disk.
As many group I sources are resolved at 25\,$\mu$m, we suggest that many, not all, group I Herbig Ae/Be disks have a hole or gap and are (pre-)transitional disks.
On the other hand, the unresolved nature of many group II sources at 25\,$\mu$m
supports that group II disks have continuous flat disk geometry.
It has been inferred that group I disks may evolve into group II through settling of dust grains to the mid-plane of the proto-planetary disk.
However, considering growing evidence for the presence of a hole or gaps in the disk of group I sources, such an evolutionary scenario is unlikely. 
The difference between groups I and II may reflect different evolutionary pathways of protoplanetary disks.
\end{abstract}

\section{Introduction}
Recent discoveries of numerous exoplanets have been revealing the diversity of planetary systems \cite[e.g.][]{Marois2008}. However, the origin of such variety is still uncertain.
Planets should have formed in the protoplanetary disks and it is essential to 
understand their evolution in order to resolve why such differences
exist.
In studies of lower-mass young stars such as T Tauri stars, transitional disks have received attention from a view of 
planet formation. Transitional or pre-transitional disks are protoplanetary disks with an inner hole and/or gaps 
indicated by the weak near-infrared(NIR)/mid-infrared(MIR) excess in their spectral energy distribution (SED) \citep{Strom1989,Espaillat2007}.
Since a primordial protoplanetary disk must have continuous distributions of dust/gas without gaps 
and the disk structure will be affected by planet formation,
those disks with an inner hole and/or gaps must be in a transitional phase from a primordial to an evolved planetary-system stage.

Disks around nearby Herbig Ae/Be stars have also been studied extensively 
in the context of the disk evolution and planet formation, but in a different classification approach.
Based on an analysis of SEDs, 
\cite{Meeus2001} classified Herbig Ae/Be stars into two groups: group I sources, which show both a power-law and a blackbody components up to far-infrared (FIR) wavelengths in their SEDs, and group II sources, whose SEDs can be well modeled with only a single power-law from MIR to FIR wavelengths. They suggested that group I has a flaring disk, while the disk around group II is geometrically flat.

There are several scenarios proposed for 
an evolutionary link between groups I and II sources. \cite{Dullemond2004b} showed that SEDs of
group I sources can be interpreted as hydrostatic disks with flaring geometry, while group II sources are an evolved version of 
group I sources, which have undergone grain growth and grain settling to the mid-plane of the disk. 
Such a settled disk would become a self-shadowed disk by a
puffed-up inner rim that accounts for weak FIR emission \citep{Dullemond2004a}.
\cite{Marinas2011} performed a MIR imaging survey of Herbig Ae/Be disks at 12 and 18 \,$\mu$m. They found that group I disks show more extended emission 
than those of group II and suggested that the trend can be naturally understood in terms of the difference in the disk geometry.

Recent high-spatial resolution observations at various wavelengths have revealed a more complex structure such as a hole or gaps in disks.
In particular, there is a growing evidence for the presence of a hole and/or gaps toward group I sources, such as AB Aur \citep{Lin2006,Honda2010}, 
HD 142527 \citep{Fukagawa2006,Fujiwara2006,Verhoeff2011}, HD 135344B \citep{Brown2009,Grady2009}, HD 36112 \citep{Isella2010}, 
HD 169142 \citep{Grady2007,Honda2012}, Oph IRS 48 \citep{Geers2007}, HD 100546 \citep{Bouwman2003,Benisty2010}, 
HD 139614 \citep{Matter2014}, and HD 97048 \citep{Maaskant2013}.
Recently, \cite{Honda2012} and \cite{Maaskant2013} proposed that group I sources possess a disk with a strongly depleted inner region
(i.e. a transitional disk). Such a discontinuous structure is different from the one originally proposed for group I disks.
As little/no evidence for a hole and/or gaps is reported toward group II disks and they seem to have a radially continous structure, 
previous interpretation on an evolutional path from group I to group II needs to be reconsidered.

In this paper, we present results of an imaging survey of nearby (roughly within 200 pc) 
Herbig Ae/Be stars at 24.5\,$\mu$m using the 8.2-m Subaru Telescope and 8.1-m Gemini Telscope. 
At 24.5\,$\mu$m, the point spread function (PSF) is relatively stable
compared to those at shorter wavelengths because of a larger Fried length, which
enables us to discuss small extended structures with high reliability. 
In addition, it allows us to investigate the cooler outer part of the disk at a dust temperature $\sim$100 K.
Early examples of our imaging survey have been published \citep{Fujiwara2006,Honda2010,Honda2012,Maaskant2013}. 
This paper gives a summary of the survey.


\section{Observations and Data Reduction}
\subsection{Subaru/COMICS data}

We made imaging observations of Herbig Ae/Be stars using 
COMICS \citep[Cooled Mid-Infrared
Camera and Spectrometer;][]{Kataza2000,Okamoto2003,Sako2003}
on the 8.2-m Subaru Telescope with the 
Q24.5 filter ($\lambda_c$=24.5\,$\mu$m,
$\Delta\lambda$=0.8\,$\mu$m). We also observed part of the targets with the Q18.8 filter ($\lambda_c$=18.8\,$\mu$m,
$\Delta\lambda$=0.8\,$\mu$m).
The chopping throw was 10$''$ and the chopping frequency was 0.45 Hz. The pixel scale is 0.130$''$/pix. 
Immediately before and/or after the observations of the target, we made observations of PSF reference stars. 
A summary of the observations is given in Table \ref{obssummary}.

For the data reduction, we employed a shift-and-add method 
to rectify the blurring caused by tracking and/or
registration errors.
The imaging data consist of 0.983 sec on-source
integration frames of coadded-exposures at each beam position. 
First, the fluctuation of the thermal background and the dark current
signals were removed by the differentiation of the chopped pair
frames. The object is bright enough to be recognized even in 0.983 sec
integration chop-subtracted frames. 
We estimated the peak position of the source by a Gaussian fitting without difficulty.  
Then we shifted the frames so as to
align the peak position and summed up the frames. We excluded the
frames whose Gaussian full-widths at half-maximum (FWHMs) deviate 
more than 1 $\sigma$ from the mean value.

\subsection{Gemini/T-ReCS data}

Observations were performed using T-ReCS \citep{Telesco1998} on the 
8.1-m Gemini South telescope. T-ReCS uses a Raytheon 320 $\times$ 240 pixel 
Si:As IBC array, with a pixel scale of 0.08633$\pm$0.00013$''$ pixel$^{-1}$, 
providing a field of view (FOV) of 27.6$''$ $\times$ 20.7$''$. 
The Q$_{\mbox{b}}$ ($\lambda_{c}$ = 24.56 \um, $\Delta\lambda$ = 0.94 \um, 50\% cut-on/off) 
filter was used in the present observations. A summary of the observations 
is shown also in Table \ref{obssummary}. Observations were made using a standard 
chop-nod technique to remove time-variable sky background and telescope 
thermal emission, and to reduce the effect of 1/{\it f} noise from the 
array-electronics. In all observations the chop-throw was 15$''$, 
the chop-angle 45\degree~E of N, and the telescope was nodded approximately every 40 s.
Standard stars were observed immediately before and/or after each object 
observation using the same instrumental configuration. 

The data were reduced using the Gemini IRAF package. The difference 
for each chopped pair was calculated and a pair of the nod sets were then differentiated
and combined to create a single image. 
All the nodding data were examined if they had high background due either to the presence 
of terrestrial clouds or temporarily high water vapor precipitation.  No data were found 
to be severely affected by these problems.

Although there is a slight difference in the characteristics 
of the filters used by COMICS and T-ReCS, 
we will refer to Q24.5 and Q$_b$ as 25\,$\mu$m throughout this paper.

\section{Results}
\begin{figure}
\plotone{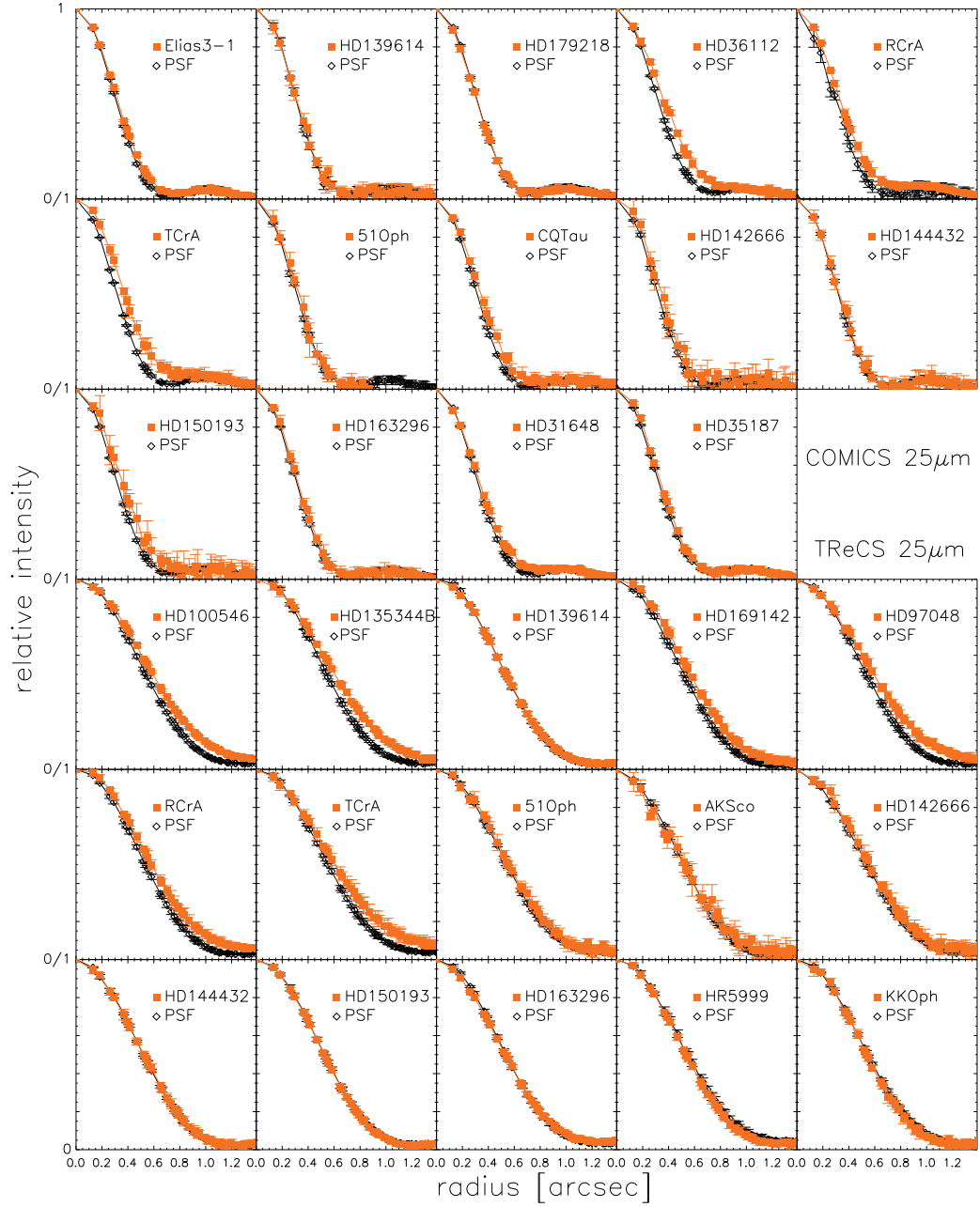}
\caption{Peak-normalized azimuthally averaged radial profile plot of the targets and PSF stars at 25\,$\mu$m.
The panels of upper 3 rows are from COMICS observations, while those of lower 3 rows are from T-ReCS observations.
\label{radiprofigure}}
\end{figure}

Since the observed images show circularly-symmetric shape and 
the azimuthal variation is not significant,
we focus on the radial profile of the targets and 
do not discuss azimuthal structures in this study.
We first made azimuthally-averaged radial profiles of the
targets and relevant PSF stars at 25\,$\mu$m as shown in
Figure \ref{radiprofigure}. We then measured their FWHMs 
from the profiles directly; we call these `direct FWHMs'
($\Phi_{d,target}$ and $\Phi_{d,PSF}$ for the targets and PSFs,
respectively).
These FWHMs are the real extension of the sources convolved with the instrumental FWHM.
These measurements are summarized in Table \ref{Table2} accompanied with those of the corresponding PSF stars.


The radial brightness profiles of most targets are comparable to or slightly wider compared to that of the PSF stars.
As a quantitative measure of the intrinsic size of the MIR emission from the disk, 
we employ a quadrature-subtraction of the FWHM of the PSF star from that of the target following \cite{Marinas2011}.
We call this `intrinsic FWHM' ($\Phi_{i}$), which is derived from  
$$\Phi_{i} = \sqrt{\Phi_{d,target}^2-\Phi_{d,PSF}^2}.$$
Although this method provides a correct size only
when the intrinsic radial profiles of both target and the PSF star are given by a Gaussian,
we adopt this method to semi-quantitatively discuss the extension of the sources with the same measure for the sake of simplicity.
Eight sources are observed by both COMICS and T-ReCS, 
and the results were consistent with each other within the measurement errors.
To be conservative, we adopt the smaller and more stringent 
value of the intrinsic FWHM on these cases.
The derived values are summarized in Table \ref{Table3}.

\section{Discussion}

\subsection{Trends on the extended emission}
To investigate possible trends of the 25\,$\mu$m extension of the Herbig Ae/Be stars with other parameters, 
we collected the distance, stellar luminosity, classification of the group proposed by \cite{Meeus2001}, 
and the MIR spectral index given by the flux density ratio at 13.5 and 30\,$\mu$m \citep{Acke2004,Acke2006,Acke2010,Meeus2012}. The flux densities at 13.5 and 30\,$\mu$m reflect the underlying continuum shape and are chosen to avoid MIR dust features such as silicates and polycyclic aromatic hydrocarbons (PAHs).
For the objects whose spectral index is not available, 
we calculated it ourselves using the {\it ISO} or {\it Spitzer} archive spectra.
We also converted the diameter in arcseconds to AU using the distance to the objects given in Table \ref{Table3}.
We added AB Aur, which was part of our survey but
its results were published earlier in \cite{Honda2010}, to Table 3.

We notice that group I sources tend to show more extended 
MIR emission than group II sources.
Nine out of 11 group I sources are resolved (i.e., 82\%)
with a signal-to-noise ratio (S/N) larger than three, and so
are 4 out of 11 group II sources, which, however, is only
36\%.
This trend is similar to the results of the study by \cite{Marinas2011} at 12 and 18\,$\mu$m.
The present results confirm the trend also at 25\,$\mu$m.

\begin{figure}
\plotone{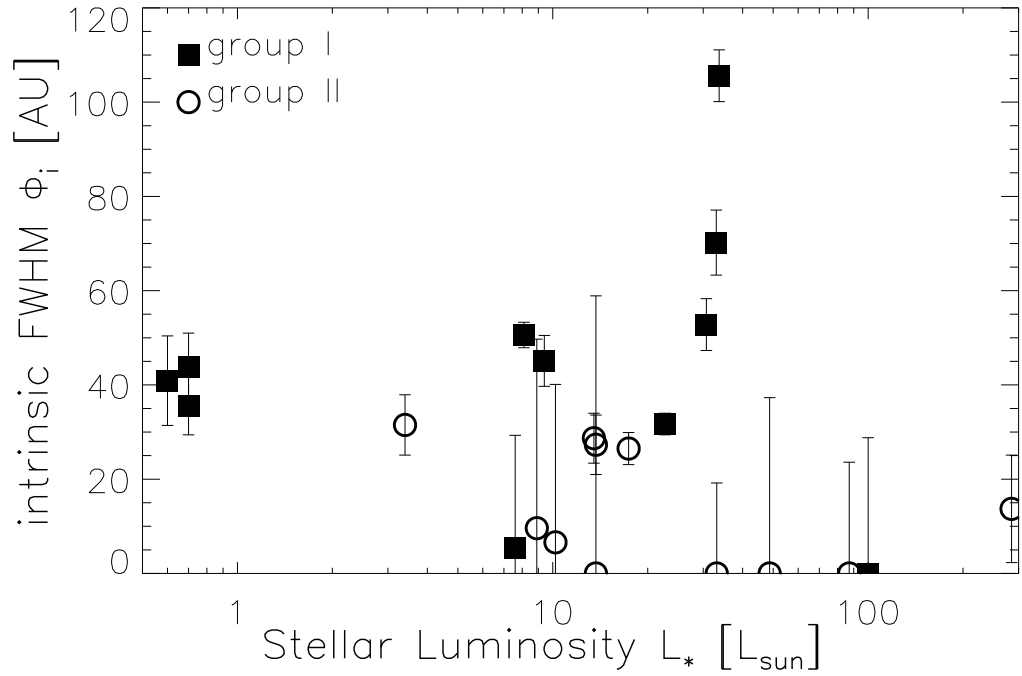}
\caption{Intrinsic FWHM of extended emission at 25\,$\mu$m against
the stellar luminosity. The squares indicate group I sources,
while the circles show group II sources. The points plotted at FWHM value of zero
are those of unresolved sources with 3-sigma upper limits.
\label{L-FWHMplot}}
\end{figure}

In Figure \ref{L-FWHMplot}, we plot the intrinsic FWHM ($\Phi_i$) against
the stellar luminosity ($L_*$).
One may expect that luminous sources show more extended emission, 
however, we could not find clear trend in the plot. 
Some sources in our sample are not resolved even though they are luminous ($L_* > 40 L_\odot$). 

\begin{figure}
\plotone{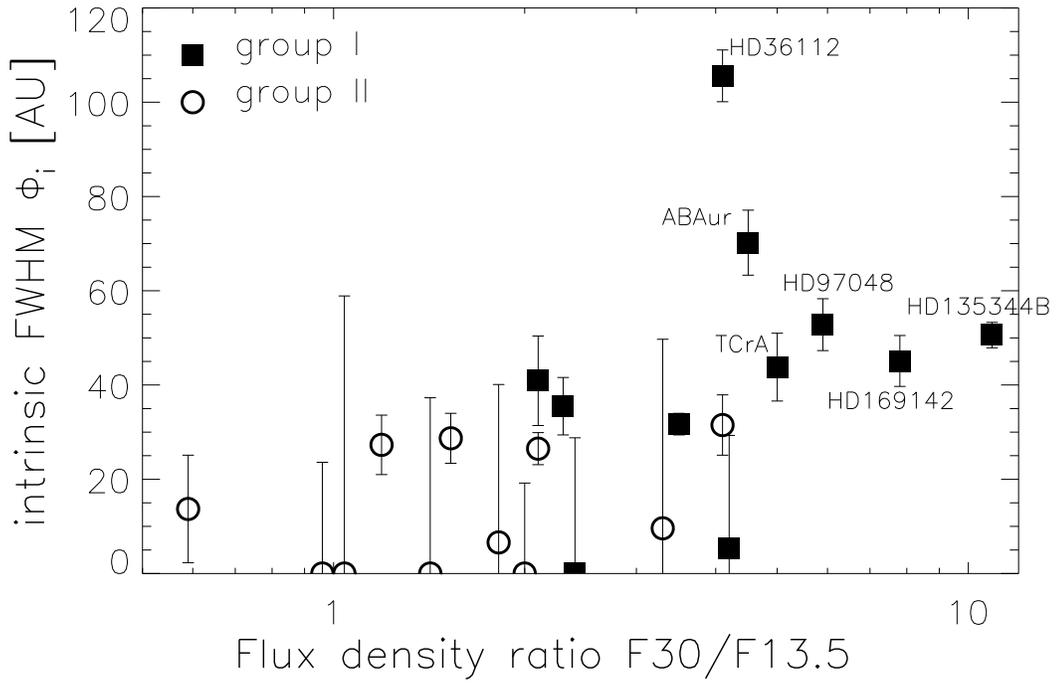}
\caption{Intrinsic FWHM of extended emission at 25\,$\mu$m against 
the MIR continuum flux ratio $F_{30}/F_{13.5}$. 
The symbols are the same as in Figure \ref{L-FWHMplot}.
Redder sources tend to show larger FWHM value.
\label{MIRcolor-FWHMplot}}
\end{figure}

On the other hand, when we plot the intrinsic FWHM
against the MIR spectral index (Figure \ref{MIRcolor-FWHMplot}), 
we find that significantly extended (FWHM $>$ 40 AU) sources all belong
to `red' group I.
Such objects exhibit the MIR spectral indices [30/13.5]
larger than 4.2, while moderately extended or unresolved
sources all show below that value, even amongst group I.
It is also interesting to note that the MIR spectral indices of well-resolved MIR disk sources such as HD142527 \citep{Fujiwara2006}, Oph IRS48 \citep{Geers2007}, and HD141569 \citep{Fisher2000,Marsh2002} are 5, 10.4, and 6.8, respectively, in accordance with the present finding.
We therefore suggest that the redder Herbig Ae/Be stars
with the MIR spectral index larger than ~4.2 exhibit more
extended MIR emission.
In general, group I sources tend to show MIR continuum emission redder than those in group II.
Thus the present finding is consistent with the trend that group I sources are likely to exhibit more extended emission than group II sources.

\subsection{Origin of extended emission of MIR red source}
The origin of {\it Q-}band (16--25\,$\mu$m) extended emission of the group I Herbig Ae/Be stars or red MIR sources have so far been discussed by several groups.
\cite{Honda2010,Honda2012} and \cite{Maaskant2013} demonstrate the difficulty in explaining the extended {\it Q-}band emission of group I sources
with a continuous disk. 
The {\it Q-}band emission from a continuous disk is mostly originated from dust grains located in the inner $\leq$ 10 AU, 
which corresponds to $\sim$0.07$''$, if located at a typical distance of $\sim$150 pc to our targets.
Considering the PSF size ($\sim$0.7$''$) at 25\,$\mu$m, this is too small to be resolved with 8 meter class telescopes.
This situation may apply to most unresolved targets in our sample.
In contrast, we have definitely resolved many group I sources, 
indicating that the continuous disk interpretation is not valid for these objects.

The shape of SEDs for group I sources can be interpreted
as having a MIR dip because of the rising FIR emission.
The dip indicates that hot/warm dust grains responsible
for the MIR radiation are depleted in the inner region of
the protoplanetary disk.
An inner hole and/or gaps in the disk can naturally
explain both the MIR dip in the SED and the extended
emission in the {\it Q-}band.
The presence of an inner hole, for example, causes the
inner edge of the disk to be directly illuminated by the
central star. This edge, being relatively further away
because of the inner hole, produces the red MIR index
(i.e, a large [30/13.5] ratio) and the strong FIR
radiation as reflected in the SED, as well as the extended
{\it Q-}band emission.
In fact, significantly extended sources (intrinsic FWHM $>$ 40AU) in our sample exhibit MIR spectral indices [30/13.5] larger than 4.2, 
indicating that the dust temperature of the inner edge of the outer disk to be $\sim$155 K assuming a blackbody.
Were the luminosity of the central star 30 $L_\odot$ 
(a typical value for well extended group I sources), 
the distance to the 155-K blackbody would be at about 18 AU,
indicating an inner hole diameter of 36 AU, which
corresponds to approximately 0.24$''$ if located at a typical
distance of 150 pc.
An emission region of this size, when convolved with the
PSF of the telescope, can be (marginally) resolved with
8-meter class telescopes in the {\it Q-}band.
We suggest that this applies to our resolved sample.

This interpretation is supported by an increasing number of detections of inner holes and gaps in group I protoplanetary disks by high-spatial resolution observations 
in the NIR and at radio wavelengths as described in section 1. 
On the other hand, little evidence has been reported for inner holes and/or gaps towards group II sources, 
which is also consistent with the present results of unresolved or limited extended emission. 
Continuous disks seem to be rather difficult to resolve in
the {\it Q-}band with 8-m class telescopes.

In general, group I sources tend to show redder MIR continuum emission than group II sources.
In our sample, most group I sources show the MIR spectral index higher than 2, while that of most group II objects is below 2.
This is consistent with a recent classification criterion put forward by \cite{Khalefinejad2014} that the MIR index [30/13.5] of group I sources is
greater than 2.1.
A blackbody at T $\sim$ 195 K would yeild a MIR index of 2.
Thus the dust temperature of the inner edge of the outer
disk around group I objects must be below 195 K, which
puts the inner edge at some distance from the central
star, producing an extended Q-band emission.
Both the high MIR index and the extended {\it Q-}band emission can naturally be accounted for by the presence of the inner edge of the outer disk located at some distance. 
As mentioned earlier, there is now growing evidence that
an inner hole and/or gaps exist in the protoplanetary
disk of group I sources.  Our finding above 
(the general trend between the extended Q-band emission and the
MIR color for group I objects) also appears to reconfirm
this view.

\subsection{Comparison with models}
 To show the effect of a gap in the disk on the Q-band image size, 
we constructed disk models and derived FWHM of the disk image at 25\,$\mu$m for comparison with the observations.
We follow the model used in \cite{Maaskant2013} who employed 
the radiative transfer tool MCMax \citep{Min2009}.
Parameters used in this model are summarized in Table \ref{Table4}.
We focus on the model with these typical Herbig Ae parameters as an example only, 
and we are not going to construct models that best predict individual imaging results.

\begin{figure}
\epsscale{1.2}
\plotone{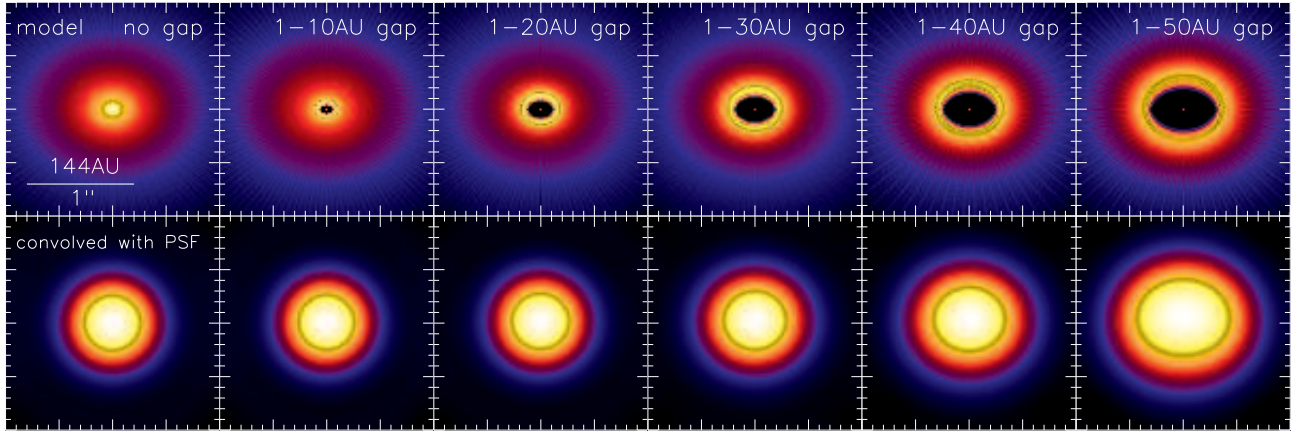}
\caption{(Top) Model images of a Herbig Ae disk with typical parameters by changing gap outer radius at 
25\,$\mu$m. Images are shown in logarithmic scale to show faint outer disk emission. 
(Bottom) Model images convolved with the PSF, assumed to be a Airy function, with a FWHM of 0.65$''$.
Images are shown in linear scale.
\label{ModelImage}}
\end{figure}

\begin{figure}
\plotone{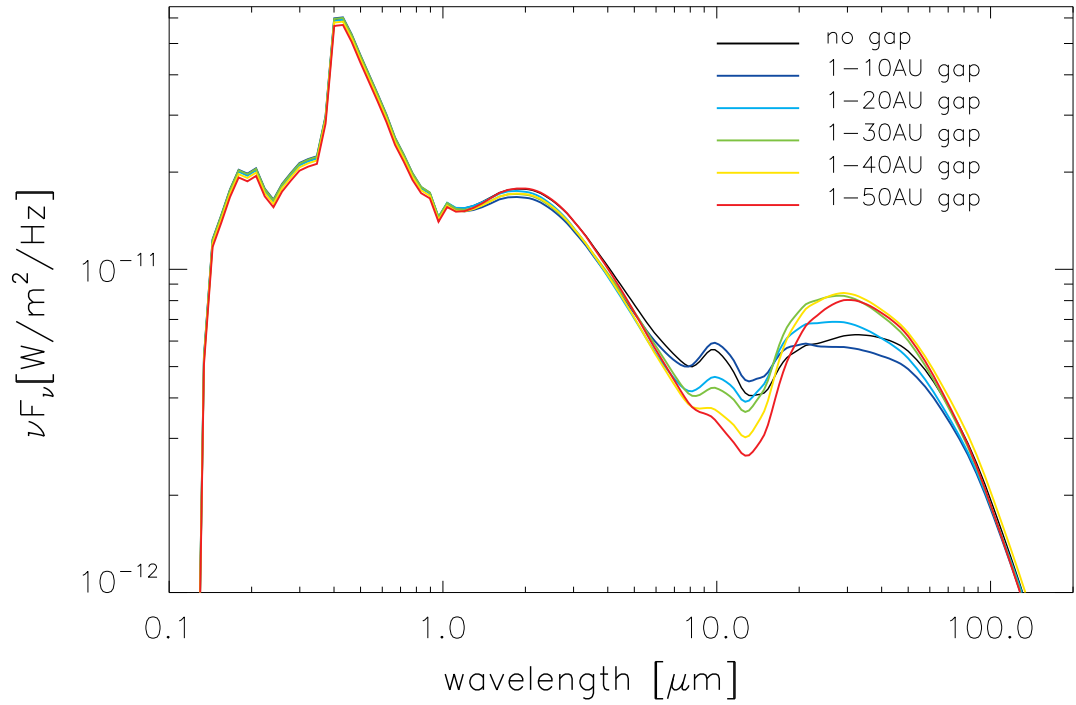}
\caption{SEDs of each models shown in Fig. \ref{ModelImage}. 
\label{SED-Model}}
\end{figure}

\begin{figure}
\plotone{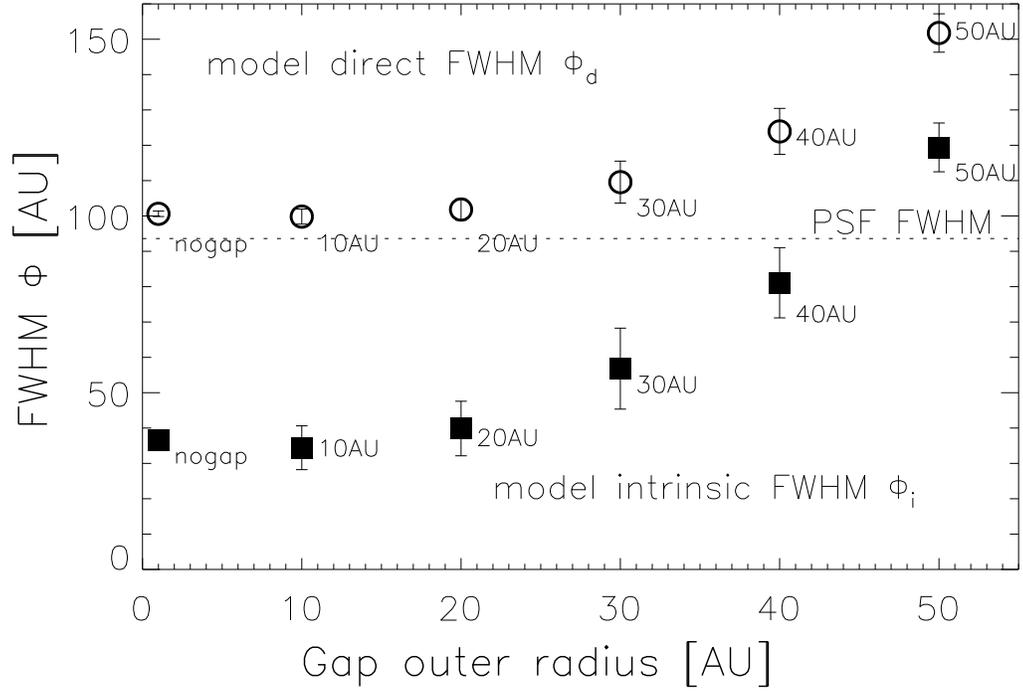}
\caption{The direct FWHM $\Phi_d$ of the PSF convolved model image shown in Fig. \ref{ModelImage} plotted against the outer gap radius.
Similar plot of instrinsic FWHM $\Phi_i$ derived via quadratic subtraction method is also shown. Disks with large gap ( gap outer radius is larger than 20 AU) show larger FWHM values than those with no or small gap.
\label{Gap-FWHMplot-Model}}
\end{figure}

\begin{figure}
\plotone{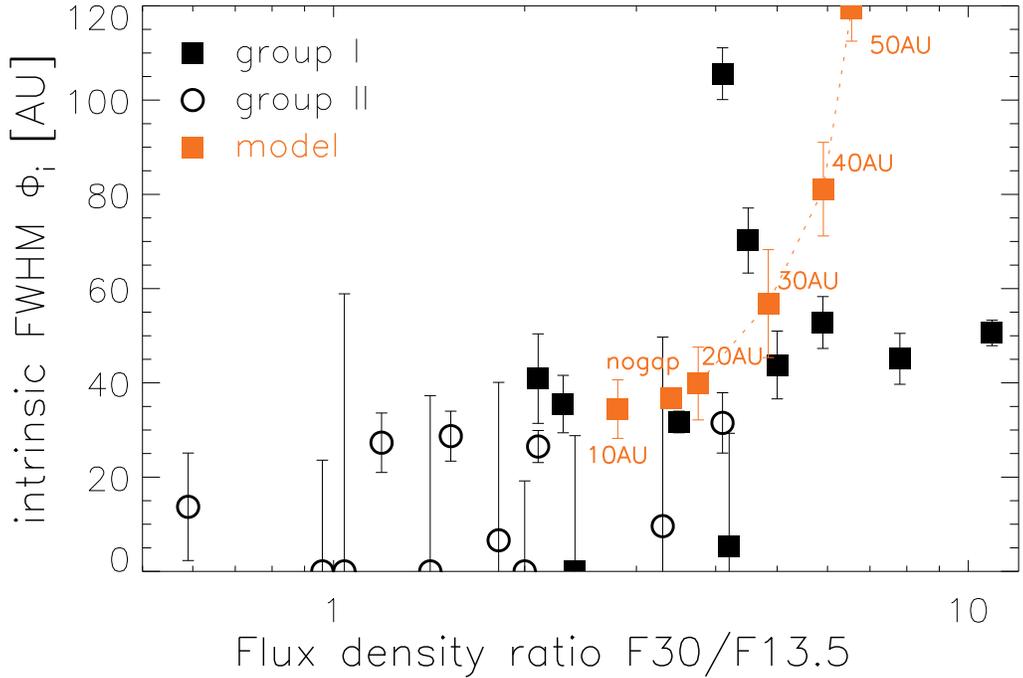}
\caption{Plot of instrinsic FWHM $\Phi_i$ of the model (red) against the MIR color overplotted in Fig.\ref{MIRcolor-FWHMplot}. 
Again, disks with large gap ( gap outer radius is larger than 20 AU) show larger MIR index [30/13.5] 
than those with no or small gap. In other words, redder sources show more extended emission.
\label{MIRcolor-FWHMplot-Model.eps}}
\end{figure}

First, we constructed a cotinuous disk model (no gap) whose 
SED has a rising FIR flux density similar to the group I objects.
The model image at 25\,$\mu$m is displayed in the top panel of Fig.\ref{ModelImage}, 
and the SED in Fig.\ref{SED-Model}.
Then we introduced a radial gap in this model. 
The gap inner radius is fixed at 1 AU, 
and the gap outer radius was increased from 10 to 50 AU in a 10-AU interval.
Images and SEDs for these models are shown in Fig.\ref{ModelImage} and Fig.\ref{SED-Model}, respectively.
The morphology of the gapped disk images is dominated by
a ring-like emission arising from the inner edge of the
outer disk (see top panels of Fig.\ref{ModelImage}).
As the size of the gap is increased, more thermal
radiation from the increasingly larger gapped area is
removed from the SED, resulting in a weakened MIR
emission and an enhanced FIR flux (see Fig.\ref{SED-Model}).
This in turn was reflected in a growingly redder
[30/13.5] colour as the gap size widened.
The disk images were subsequently convolved with an
Airy pattern with a FWHM = 0.65$''$ (assumed to
represent the telescope beam; see bottom panels of
Fig.\ref{ModelImage}).
We measured $\Phi_d$ and derived $\Phi_i$ from these final
images in the same manner as described in Section 3 (see
Fig.\ref{Gap-FWHMplot-Model}).
As can be seen, the models with outer gap radii smaller
than 20 AU are comparable with the `nogap' model
(equivalent within uncertainties), implying that
they would either be unresolved or only marginally
resolved at best.
On the other hand, when the outer gap radius $\ga 30$ AU,
they would easily be resolved at 25\,$\mu$m.
Thus our 25\,$\mu$m imaging survey is sensitive to the presence of a large ($\ga 30$~AU) gap.
Fig.\ref{MIRcolor-FWHMplot-Model.eps} is the same as Fig.\ref{MIRcolor-FWHMplot}, 
except we added the models points.
In our sample models, disks with [30/13.5] $\ga$ 4.2 can be achieved 
when the gap outer radius becomes $\ga 25$~AU and
such disk can be well-resolved in our observations.
This trend is almost consistent with our observational findings 
that the well-extended Herbig Ae/Be sources show MIR index larger than 4.2.
Again, it demonstrates that our 25\,$\mu$m imaging survey is sensitive to disks with large gaps.

\subsection{Group I sources as (pre-)transitional disks}
Since the presence of an inner hole and/or gaps has been shown to be a common characteristic for group I Herbig Ae/Be stars \citep[e.g.][]{Lin2006,Fukagawa2006,Fujiwara2006,Verhoeff2011,Brown2009,Grady2007,Grady2009,Isella2010,Honda2010,Honda2012,Geers2007,Bouwman2003,Benisty2010,Matter2014,Maaskant2013},
\cite{Honda2012} and \cite{Maaskant2013} suggest that most group I sources can be classified as (pre-)transitional disks.
Transitional or pre-transitional disks, which are originally suggested for low-mass young stars such as T-Tauri stars, are protoplanetary disks with an inner hole and/or gaps indicated by the weak NIR excess in the SED \citep{Strom1989,Espaillat2007}.
Because the primordial disk is thought to have a continuous distribution of dust without gaps and because planet formation could produce a hole and/or gaps in the disk,
those disks with (large) inner hole and/or gaps must be in a transitional phase from a primordial to an evolved stage.

On the other hand, an evolutionary scenario for Herbig Ae/Be stars is still a matter of debate. 
\cite{Meeus2001} proposed that group I disk is flared while that of group II is flat, based on the analysis of the SED.
A possible evolutionary scenario was suggested in which group I flaring disk evolves into group II flat disk by grain growth 
and sedimentation/settling of grains to the disk mid-plane. 
However, the present study indicates that the group I disk is a (pre-)transitional disk with an inner cleared region and/or gaps,
while the group II disk is a continuous disk.
These observational pieces of information imply that evolution from group I into group II is unlikely.
As \cite{Meeus2001} pointed out that there is no significant difference of the age between groups I and II, 
it is more likely that both sources may have evolved differently from a primordial continuous flaring disk, a common ancester as discussed in \cite{Maaskant2013}.
This scenario is quite similar to the T Tauri disk evolutionary scenario proposed by \cite{Currie2010}.
He presented two main pathways for the evolution of T Tauri disks: those that form an inner hole/gap and others that deplete more homologously.
The present study 
suggests a similarity between the evolutionary scenarios of T Tauri and Herbig Ae/Be disks.


\acknowledgments
We are grateful to all of the staff members of the Subaru Telescope.
We also thank Dr. Hitomi Kobayashi and Dr. Yuji Ikeda at Kyoto-Nijikoubou Co., Ltd..
This research was partially supported by KAKENHI (Grant-in-Aid for Young
Scientists B: 21740141) by the Ministry of Education, Culture, Sports,
Science and Technology (MEXT) of Japan.


\begin{deluxetable}{lccccccccc}
\tablecaption{Summary of Subaru/COMICS and Gemini/T-ReCS Observations\label{obssummary}}
\tablehead{
& \multicolumn{3}{c}{Subaru/COMICS Q24.5 imaging} &
\multicolumn{3}{c}{Subaru/COMICS Q18.8 imaging}   &
\multicolumn{3}{c}{Gemini/T-ReCS Qb imaging}   \\
\tableline
\colhead{Object} & \colhead{Date} & \colhead{t\tablenotemark{a}} & \colhead{PSF} &
\colhead{Date} & \colhead{t\tablenotemark{a}} & \colhead{PSF} &
\colhead{Date} & \colhead{t\tablenotemark{a}} & \colhead{PSF} }
\startdata
Elias3-1 & Jul 11, 2004, Jul 12, 2004 & 399 & $\beta$And & ... & ... & ... & ...          & ... & ... \\
HD100546 & ...                        & ... & ...        & ... & ... & ... & Jun 27, 2011 & 638 & $\gamma$Cru \\
HD135344B& ...                        & ... & ...        & ... & ... & ... & Jun 14, 2011 & 1361 & $\alpha$Cen A \\
HD139614 & Jul 11, 2004               & 101 & $\delta$Oph & ...& ... & ... & Jul 23, 2011 & 638 & $\alpha$ Tra\\
HD169142 & ...                        & ... & ...        & ... & ... & ... & Jul 22, 2011 & 638 & $\eta$ Sgr \\
HD179218 & Jul 11, 2004               & 99  & $\alpha$Her &... & ... & ... & ...          & ... & ... \\
HD36112  & Dec 14, 2005, Jan 26, 2011 & 3297 & $\alpha$Tau& Jan 26, 2011 & 438 & $\alpha$Tau & ... & ... & ... \\
HD97048  & ...                        & ... & ...        & ... & ... & ... & Jun 28, 2011 & 638 & $\gamma$Cru	\\
RCrA     & Jul 11, 2004               & 100 & $\delta$Oph & Jul 12, 2004 & 40 & $\alpha$Her & Jun 28, 2011 & 203 & $\eta$ Sgr \\
TCrA     & Jul 11, 2004               & 312 & $\alpha$Her & Jul 12, 2004 & 175 & $\alpha$Her & Jul 21, 2011 & 638 & $\eta$ Sgr \\
51 Oph   & Jul 11, 2004               & 237 & $\delta$Oph & ... & ... & ... & Jul 21, 2011 &638 & $\eta$ Sgr \\
AK Sco   & ...                        & ... & ...         & ... & ... & ... & Jul 22, 2011 &638 & $\eta$ Sgr \\
CQTau    & Dec 15, 2005               & 553 & $\alpha$Tau & Dec 15, 2005 & 541 & $\alpha$Tau & ... & ... & ... \\
HD142666 & Jul 11, 2004               & 148 & $\delta$Oph & ... & ... & ... & Jul 21, 2011 & 638 & $\eta$ Sgr \\
HD144432 & Jul 11, 2004               & 193 & $\delta$Oph & ... & ... & ... & Jul 24, 2011 & 638 & $\delta$ Oph\\
HD150193 & Jul 12, 2004               & 168 & $\alpha$Her & ... & ... & ... & Jul 27, 2011 & 638 & $\eta$ Sgr \\
HD163296 & Jul 11, 2004, Jul 12, 2004 & 557 & $\delta$Oph & ... & ... & ... & Jun 28, 2011 & 638 & $\eta$ Sgr \\
HD31648  & Dec 14, 2005               & 1510 & $\alpha$Tau & Dec 15, 2005 & 578 & $\alpha$Tau & ... & ... & ... \\
HD35187  & Dec 14, 2005, Dec 16, 2005 & 1580 & $\alpha$Tau & ...          & ... & ...         & ... & ... & ... \\
HR5999	 & ...                        & ... & ...         & ... & ... & ... & Jul 24, 2011 & 638 & $\alpha$ Tra	\\
KK Oph   & ...                        & ... & ...         & ... & ... & ... & Jul 21, 2011 & 638 & $\eta$ Sgr
\enddata
\tablenotetext{a}{Total integration time in seconds used in this study}
\end{deluxetable}


\begin{deluxetable}{lccccccccc}
\tablecaption{FWHM measurements of COMICS and T-ReCS observations\label{Table2}}
\tablehead{ & \multicolumn{3}{c}{Subaru/COMICS Q24.5 imaging} & \multicolumn{3}{c}{Subaru/COMICS Q18.8 imaging} & \multicolumn{3}{c}{Gemini/T-ReCS Qb imaging} \\
\colhead{object} & \colhead{$\Phi_{d,target}('')$} & \colhead{$\Phi_{d,PSF}('')$} & \colhead{?\tablenotemark{a}} & \colhead{$\Phi_{d,target}('')$} & \colhead{$\Phi_{d,PSF}('')$} & \colhead{?\tablenotemark{a}} & \colhead{$\Phi_{d,target}('')$} & \colhead{$\Phi_{d,PSF}('')$} & \colhead{?\tablenotemark{a}} }
\startdata
\tableline
Elias3-1  & 0.672$\pm$0.012 & 0.634$\pm$0.003 & Y  & ... & ... & ... & ...             & ...             & ... \\
HD100546  & ...             & ...             & ...& ... & ... & ... & 0.788$\pm$0.006 & 0.716$\pm$0.008 & Y \\
HD135344B & ...             & ...             & ...& ... & ... & ... & 0.804$\pm$0.008 & 0.721$\pm$0.004 & Y \\
HD139614  & 0.643$\pm$0.031 & 0.633$\pm$0.011 & N  & ... & ... & ... & 0.711$\pm$0.005 & 0.710$\pm$0.007 & N \\
HD169142  & ...             & ...             & ...& ... & ... & ... & 0.759$\pm$0.014 & 0.692$\pm$0.007 & Y \\
HD179218  & 0.637$\pm$0.009 & 0.645$\pm$0.004 & N  & ... & ... & ... & ...             & ...             & ...\\
HD36112   & 0.751$\pm$0.009 & 0.649$\pm$0.003 & Y  & 0.559$\pm$0.017 & 0.526$\pm$0.027 & N & ... & ... & ... \\
HD97048   & ...             & ...             & ...& ... & ... & ... & 0.788$\pm$0.014 & 0.714$\pm$0.005 & Y \\
RCrA      & 0.687$\pm$0.016 & 0.629$\pm$0.020 & Y  & 0.547$\pm$0.011 & 0.489$\pm$0.008 & Y & 0.771$\pm$0.028 & 0.704$\pm$0.011 & Y \\
TCrA      & 0.748$\pm$0.013 & 0.634$\pm$0.002 & Y  & 0.586$\pm$0.026 & 0.489$\pm$0.004 & Y & 0.806$\pm$0.023 & 0.732$\pm$0.003 & Y \\

51 Oph   & 0.663$\pm$0.037 & 0.626$\pm$0.006 & N   & ... & ... & ... & 0.730$\pm$0.012 & 0.721$\pm$0.007 & N \\
AK Sco   & ...             & ...             & ... & ... & ... & ... & 0.695$\pm$0.024 & 0.692$\pm$0.005 & N \\
CQTau    & 0.687$\pm$0.023 & 0.627$\pm$0.004 & Y & 0.519$\pm$0.005 & 0.501$\pm$0.011 & Y? & ... & ... & ... \\
HD142666 & 0.710$\pm$0.063 & 0.637$\pm$0.008 & N & ...             & ...             & ...& 0.741$\pm$0.008 & 0.714$\pm$0.005 & Y \\
HD144432 & 0.652$\pm$0.031 & 0.639$\pm$0.006 & N & ...             & ...             & ...& 0.691$\pm$0.014 & 0.690$\pm$0.005 & N \\
HD150193 & 0.728$\pm$0.081 & 0.641$\pm$0.002 & N & ...             & ...             & ...& 0.696$\pm$0.008 & 0.700$\pm$0.007 & N \\
HD163296 & 0.649$\pm$0.011 & 0.632$\pm$0.003 & Y?& ...             & ...             & ...& 0.705$\pm$0.008 & 0.711$\pm$0.006 & N \\
HD31648  & 0.677$\pm$0.013 & 0.646$\pm$0.006 & Y & 0.503$\pm$0.007 & 0.493$\pm$0.007 & N  & ...             & ...             & ...\\
HD35187  & 0.689$\pm$0.010 & 0.649$\pm$0.002 & Y & ...             & ...             & ...& ...             & ...             & ...\\
HR5999   & ...             & ...             & ... & ...           & ...             & ...& 0.709$\pm$0.009 & 0.728$\pm$0.012 & N \\
KK Oph   & ...             & ...             & ... & ...           & ...             & ...& 0.721$\pm$0.012 & 0.722$\pm$0.007 & N 
\enddata
\tablenotetext{a}{Resolved(Y) or not(N)}

\end{deluxetable}


\begin{deluxetable}{lccccccccc}
\tablecaption{Summary of parameters of the samples\label{Table3}.
}
\tablehead{
\colhead{Object} & \colhead{Distance(pc)} & \colhead{L$_*$($L_\odot$)} & \colhead{Ref.} & \colhead{$\Phi_{i}('')$}
& \colhead{$\Phi_{i}(AU)$}  & \colhead{Group} & \colhead{Ref.} & \colhead{[30/13.5]} & \colhead{Ref.}
}
\startdata
ABAur     & 139.3 & 33.0 & b & 0.50$\pm$0.05 & 70.2$\pm$6.91 & I & a & 4.5 & a \\
Elias3-1  & 160   & 0.7  & d & 0.22$\pm$0.04 & 35.5$\pm$6.1  & I & c & 2.3 & e \\
HD100546  &  96.9 & 22.7 & b & 0.33$\pm$0.02 & 31.7$\pm$2.3  & I & a & 3.5 & a \\
HD135344B & 142   & 8.1  & b & 0.36$\pm$0.02 & 50.6$\pm$2.7  & I & a & 10.9 & a \\
HD139614  & 140   & 7.6  & b & 0.04$\pm$0.17 & 5.3$\pm$24.0  & I & a & 4.2 & a \\
HD169142  & 145   & 9.4  & b & 0.31$\pm$0.04 & 45.1$\pm$5.4  & I & a & 7.8 & a \\
HD179218  & 240   & 100.0& d & $<$0.12         & $<$28.81    & I & a & 2.4 & a \\
HD36112   & 279.3 & 33.7 & b & 0.38$\pm$0.02 & 105.6$\pm$5.5 & I & a & 4.1 & a \\
HD97048   & 158.5 & 30.7 & b & 0.33$\pm$0.03 & 52.8$\pm$5.5  & I & a & 5.9 & a \\
R CrA     & 130   & 0.6  & d & 0.31$\pm$0.07 & 40.9$\pm$9.5  & I & c & 2.1 & e \\
T CrA     & 130   & 0.7  & d & 0.34$\pm$0.06 & 43.8$\pm$7.2  & I & a & 5   & a \\

51Oph     & 124.4 & 285.0 & b & 0.11$\pm$0.09 & 13.7$\pm$11.4 & II & a & 0.59 & a \\
AK Sco    & 150   & 8.9   & d & 0.06$\pm$0.27 & 9.6$\pm$40.1  & II & a & 3.3  & a \\
CQTau     & 113   & 3.4   & b & 0.28$\pm$0.06 & 31.5$\pm$6.4  & II & b & 4.1 & e \\ 
HD142666  & 145   & 13.5  & b & 0.20$\pm$0.04 & 28.7$\pm$5.3  & II & a & 1.53 & a \\
HD144432  & 145   & 10.2  & d & 0.05$\pm$0.23 & 6.6$\pm$33.5  & II & a & 1.82 & a \\
HD150193  & 216.5 & 48.7  & b & $<$0.17       & $<$37.3       & II & a & 1.42 & a \\
HD163296  & 118.6 & 33.1  & b & $<$0.16       & $<$19.2       & II & a & 2    & a \\
HD31648   & 137   & 13.7  & b & 0.20$\pm$0.05 & 27.3$\pm$6.3  & II & a & 1.19 & a \\
HD35187   & 114.2 & 17.4  & b & 0.23$\pm$0.03 & 26.5$\pm$3.4  & II & a & 2.1  & a \\
HR5999    & 210   & 87.1  & d & $<$0.11       & $<$23.6       & II & a & 0.96 & a \\
KK Oph    & 260   & 13.7  & b & $<$0.23       & $<$58.9       & II & a & 1.04 & a
\enddata
\tablerefs{(a) \citealt{Acke2010} (b) \citealt{Meeus2012} (c) \citealt{Acke2006} (d) \citealt{Acke2004}
(e) derived from the archival spectra 
}
\end{deluxetable}


\begin{deluxetable}{lc}
\tablecaption{Parameters used in the model \label{Table4}
}
\tablehead{
\colhead{parameter} & \colhead{values} 
}
\startdata
stellar paramters :                 & \\
stellar effective temperature $T_*$ &  9280 K         \\
stellar mass $M_*$                  &  2.4 $M_\odot$  \\
stellar luminosity $L_*$            &  40. $L_\odot$ \\
distance d                          & 144. pc\\ 
\hline
disk component :                    & \\
dust composition                    & 80\% silicates + 20\% carbon \\
dust minimum size $a_{min}$           & 0.1\,$\mu$m \\
dust minimum size $a_{max}$           & 1. mm \\
index of dust size power law       & 3.75 \\
dust mass                           & 5$\times 10^{-5} M_\odot$\\
disk inner radius $r_{in}$          & 0.3 AU \\
disk outer radius $r_{out}$         & 400. AU \\
disk inclination                    & 35$^\circ$ \\
\hline
halo component :                    & \\
dust composition                    & 100\% carbon \\
dust minimum size $a_{min}$           & 0.1\,$\mu$m \\
dust minimum size $a_{max}$           & 1.\,$\mu$m \\
index of dust size power law       & 3.5 \\
dust mass                           & 1.0$\times 10^{-12} M_\odot$\\
halo inner radius $r_{in}$          & 0.25 AU \\
halo outer radius $r_{out}$         & 0.35 AU 
\enddata
\end{deluxetable}


\end{document}